\documentclass[preprint,showpacs,preprintnumbers,amsmath,amssymb]{revtex4}

\usepackage{graphicx,amsfonts}
\usepackage{epsfig}
\usepackage{dcolumn}
\usepackage{bm}
\begin{document}

\title{Multi-Higgs-doublet model with $A_4$ symmetry}

\author{A. C. B. Machado}%
\email{ana@ift.unesp.br}
\affiliation{
Instituto de F\'\i sica Te\'orica,
Universidade Estadual Paulista\\ Rua Pamplona, 145
01405-900 - S\~ao Paulo, SP, Brazil
}
\author{J. C. Montero}%
\email{montero@ift.unesp.br}
\affiliation{
Instituto de F\'\i sica Te\'orica,
Universidade Estadual Paulista\\ Rua Pamplona, 145
01405-900 - S\~ao Paulo, SP, Brazil
}
\author{V. Pleitez}%
\email{vicente@ift.unesp.br}
\affiliation{
Instituto de F\'\i sica Te\'orica,
Universidade Estadual Paulista\\ Rua Pamplona, 145
01405-900 - S\~ao Paulo, SP, Brazil
}

\date{\today}
%
\begin{abstract}
We worked out in detail the three-Higgs-doublet extension of the standard model when the $A_4$ symmetry,
which is imposed to solve the flavor problem, is extended to the scalar sector.
The three doublets may be related to the fermion mass generation and, in particular, they may be the unique responsible
for generating the neutrino masses. If this is the case, the respective VEVs have to be quite smaller  than the
electroweak scale. The usual hierarchy in the Yukawa couplings is now moved to a hierarchy in the VEVs which
may be justified on dynamical grounds. We consider here the mass spectra in the scalar sector in several situations.
When the three doublets are those related with neutrino masses some light scalar arise. However,
the later fields are safe, from the phenomenological point of view, since they couple mainly to neutrinos.

\end{abstract}


\maketitle

\section{Introduction}
\label{sec:intro}
\medskip
In the standard model of particle interactions (SM) with only one Higgs scalar doublet the fermion mass matrices remain
arbitrary. In order to become more predictive about the origin of masses and mixing, extra symmetries, continuous \cite{continuas},
or discrete~\cite{discretas}, may be imposed. On one hand, among the most motivated extensions of the SM are
those with several Higgs doublets~\cite{higgshunter}. On the other hand, the most general scalar potential even
for the simplest case of two doublets is hard to be worked out in detail~\cite{adler,turma}.
Hence, in order to study multi-Higgs extensions of the SM it is necessary to impose extra symmetries. An interesting
possibility is that these extra symmetries are the same that are also useful
for explaining the fermion masses and mixing, avoiding that the Yukawa couplings do not span
over two or even five decade of magnitude. Moreover, since all the neutral components of doublets
getting a non-zero VEV contribute to the $W^\pm$ and $Z$ masses, the sum of the VEVs squares is equal to
$\sim(174\, \textrm{GeV})^2$ and some VEVs may be smaller than 174 GeV.
Neutral Higgs scalars getting small VEVs (i.e., smaller than the electroweak scale) are
potentially dangerous since they may imply the existence of light charged or neutral (pseudo)scalars in the model.
However, the appearance of these sort of VEVs is almost inevitable when more than two doublets
are added to the SM. This situation arises in models in which different Higgs sectors
give mass to different charged fermion sectors, as in \cite{dim5,porto}.
Thus, it is necessary to search for a mechanism that allows heavy enough scalars or their phenomenological effects be
appropriately suppressed.

\section{Three doublet model with soft $A_4$ violating terms}
\label{sec:dois}
\medskip
Here we will show several vacuum alignments and see in what cases when some, or all, VEVs are small light scalar may be avoided.
This is the inverse of the mechanism considered in Ref.~\cite{duong}. We consider the case when three $SU(2)$ doublets
$h_i,\,i=1,2,3$, are in an $A_4$ triplet $h\equiv (h_1,h_2,h_3)$. The most general scalar potential invariant under $A_4$ and
the standard model gauge symmetries is denoted by $V(h)$ and we may, or not, add terms which break the $A_4$ symmetry softly,
$V_{soft}$. The scalar potential is given by
\begin{equation}
V=V(h)+V_{soft}(h_1,h_2,h_3),
\label{potential}
\end{equation}
where
\begin{eqnarray}
V(h)&=&\mu^2[h^\dagger h]_1+\lambda_1([h^\dagger h]_1)^2+\lambda_2
[h^\dagger h]_{1^\prime} [h^\dagger
h]_{1^{\prime\prime}} +\lambda_3[h^\dagger h]_A[h^\dagger h]_A
\nonumber\\&& +\lambda^*_3[h^\dagger h]_B[h^\dagger
h]_B+\lambda_4 [h^\dagger h]_A[h^\dagger h]_B,
\label{potential1}
\end{eqnarray}
which can be written explicitly in terms of $h_i$ as~\cite{ishimori}
\begin{eqnarray}
V(h)&=&\mu^2\sum_ih^\dagger_ih_i+(\lambda_1+\lambda_2)\sum_{i\not=j}(h^\dagger_ih_i)(h^\dagger_jh_j)+
\left(\lambda_3\sum_{i\not=j}(h^\dagger_ih_j)^2+H.c.\right)\nonumber \\ &+&\lambda_4\sum_{i\not=j}(h^\dagger_ih_j)(h^\dagger_jh_i),
\label{fim}
\end{eqnarray}
which was considered in \cite{lavoura}. We add also the terms that violate softly the $A_4$ symmetry:
\begin{eqnarray}
V_{soft}(h_1,h_2,h_3)&=&\mu^2_1 h^\dag_1 h_1 + \mu^2_2 h^\dag_2 h_2 + \mu^2_3 h^\dag_3 h_3
 +(\nu^2_{12} h^\dag_1 h_2 \nonumber \\&& + \nu^2_{13} h^\dag_1 h_3 + \nu^2_{23} h^\dag_2 h_3+H.c).
\label{potentialsoft}
\end{eqnarray}

We will consider, for the sake of simplicity, that all VEVs and dimensionless parameters are real  and use the
$A_4$ identity $1+\omega+\omega^2=0$. Defining $\partial V/\partial v_i=t_i$, we have
\begin{eqnarray}
&& t_1= v_1[ \mu^2_1+\mu^2  + \lambda v^2_1 + \frac{1}{2}\lambda^\prime (v^2_2 + v_3^2)]
+\nu^2_{12} v_2 + \nu^2_{13} v_3,\nonumber \\ &&
t_2= v_2[\mu^2_2+\mu^2 +  \lambda v_2^2 + \frac{1}{2}\lambda^\prime (v_1^2+v^2_3)]
+ \nu^2_{12} v_1 + \nu^2_{23} v_3,  \nonumber \\ &&
t_3= v_3[ \mu^2_3+\mu^2 + \lambda   v_3^2 + \frac{1}{2}\lambda^\prime (v_1^2+v^2_2)] + \nu^2_{13} v_1 + \nu^2_{23} v_2,
\label{minimo}
\end{eqnarray}
where $\lambda=\lambda_1+\lambda_2$, $\lambda^\prime=
2\lambda_1-\lambda_2\!\!+\lambda^{\prime\prime}$, and $\lambda^{\prime\prime}=2\lambda_3+\lambda_4$. The
constraint equations, which minimize the scalar potential, are $t_i=0$, for fixed $i$,
can be solved in several ways as we will show below.
Notice that all $v_i$ may be different from zero but may be arbitrarily small.

\subsection{First case: model with some $\mu^2_i>0$}
\label{subsec:caso1}

An example of this case is when $\mu^2<0$ and $\mu^2_1<0$ while $\mu^2_{2,3}>0$. We can obtain hierarchies among the VEVs.
Let us suppose that one of the VEVs ($v_1$) is larger than the other two ($v_2,v_3$) and that the products
$\nu^2_{12}v_2,\nu^2_{13}v_3$ are negligible compared to the first term of $t_1$,
and also that $\mu^2_2+\mu^2+\lambda v^2_2+(\lambda^\prime/2)[v^2_1+v^2_3]=0$,
$\mu^2_3+\mu^2+\lambda v^2_3+(\lambda^\prime/2)[v^2_1+v^2_2]=0$, we have from (\ref{minimo})
\begin{equation}
v_1\approx\sqrt{-\frac{\mu^2+\mu^2_1}{\lambda}},\;v_2\approx-\frac{\nu^2_{12}}{\nu^2_{23}}\,v_1,\;
v_3\approx-\frac{\nu^2_{13}}{\nu^2_{23}}\,v_1,
\label{caso1a}
\end{equation}
with $\nu^2_{12},\nu^2_{13}<0$ and $\nu^2_ {23}\gg \vert\nu^2_{12}\vert,\vert\nu^2_{13}\vert$ we have consistence with the condition
$v_2/v_1, v_3/v_1\ll1$.

On the other hand if still $v_1\gg v_2,v_3$, with $\mu^2_{2,3}$ positive but now arbitrary
and $\mu^2_{2,3}> \vert\mu^2\vert,v^2_1\gg \vert \nu^2_{12}\vert,\vert\nu^2_{13}\vert$,
we have that $v_1$ as in (\ref{caso1a}) but now
\begin{equation}
v_2\approx-\frac{2\nu^2_{12}}{2\mu^2+\lambda^\prime
v^2_1+2\mu^2_2}\,v_1,\;\;
v_3\approx- \frac{2\nu^2_{13}}{2\mu^2+\lambda^\prime
v^2_1+2\mu^2_3}\,v_1.
\label{sol1}
\end{equation}
Under these circumstances we have always a hierarchy in the VEVs as required in the model of \cite{dim5}:
$v_1\gg v_2\gg v_3$.
If the triplet that is being considered contributes to the generation of the $u-$type quark masses,
the mass spectrum of the neutral scalar sector has a standard-model-like scalar which mass is of the order of
$v_1$,  and two heavy scalars which masses are dominated by $\mu_2$ and $\mu_3$. Again, the solutions (\ref{sol1})
are consistent with the assumed condition that $v_{2,3}$ are smaller than $v_1$.
Hence, assuming the $\vert \nu^2_{12} \vert$ and $\vert \nu^2_{13} \vert $
smaller than $\vert \mu^2_2 \vert$ and $\vert \mu^2_3 \vert $,
this model is a three doublet generalization of the two doublet model of Ref.~\cite{ma01}.

\subsection{Second case: model with  all $\mu^2_i<0$}
\label{subsec:caso2}

Now, the equations (\ref{minimo}) are solved by
using the conditions $\mu^2_{1,2,3}<0$ and the resulting scalar mass spectra are as follows.
After using the conditions $t_i=0$ in Eq.~(\ref{minimo}), we obtain the mass matrices of each charge sector.
For the neutral pseudoscalar sector, in the basis $(a_1,a_2,a_3)$,  the mass matrix is
\begin{equation}
M^2_a=\left(\begin{array}{ccc}
-\frac{2\lambda_3v_1(v^2_2+v^2_3)+v_2\nu^2_{12}+v_3\nu^2_{13}}{v_1}   &
\;\;2\lambda_3 v_1v_2+\nu^2_{12} &\;\; 2\lambda_3 v_1v_3+\nu^2_{13}\\
 &  -\frac{2\lambda_3v_2(v^2_1+v^2_3)+v_1\nu^2_{12}+v_3\nu^2_{23}}{v_2}
 & \;\;2\lambda_3 v_2v_3 +\nu_{23}^2\\
 &  & -\frac{2\lambda_3v_3(v^2_1+v^2_2)+v_1\nu^2_{13}+v_2\nu^2_{23}}{v_3}
\end{array}\right).
\label{mpseudo}
\end{equation}
We will denote the eigenvectors of the matrix in Eq.~(\ref{mpseudo}) as $G^0$ for the would be
Goldstone boson, and $A^0_{1,2}$ for the two physical pseudoscalars. In fact, the mass matrix in
Eq.~(\ref{mpseudo}) has $\textrm{Det} M^2_a=0$. Here and below we will not show the eigenvectors explicitly.

In the neutral scalar sector, in the basis $(h^0_1,h^0_2,h^0_3)$, we have the mass matrix
\begin{equation}
M^2_s=\left(\begin{array}{ccc}
\frac{2\lambda v^3_1-v_2\nu^2_{12}-v_3\nu^2_{13}}{v_1}  \,\, &
\lambda^\prime v_1v_2+\nu_{12}^2  &\,\,
\lambda^\prime  v_1v_3+\nu^2_{13}\\
& \frac{2\lambda v^3_2-v_1\nu^2_{12}-v_3\nu^2_{23}}{v_2} & \lambda^\prime  v_2v_3+\nu_{23}^2 \\
&  & \frac{2\lambda v^3_3-v_1\nu^2_{13}-v_2\nu^2_{23}}{v_3}
\end{array}\right).
\label{mfisicos}
\end{equation}
The respective eigenvectors of $M^2_s$ are denoted by $S^0_{1,2,3}$, and one
of them may or may not correspond to the SM Higgs field. From (\ref{mfisicos}) we obtain $\textrm{Det}M^2_s\not=0$ hence
this sector has no Goldstone bosons as it must be. However, it is not obvious if there exist or not a light scalar
when all VEVs, but not the $\nu$s, are small.

For the charged scalar sector we have, in the basis $(h^+_1,h^+_2,h^+_3)$,
\begin{equation}
M^2_c=\left(\begin{array}{ccc}
-\frac{\lambda^{\prime\prime}v_1(v^2_2+v^2_3)+2(v_2\nu^2_{12}+v_3\nu^2_{13})}{2v_1}&\;\;
\frac{1}{2}\lambda^{\prime\prime} v_1 v_2+\nu_{12}^2  &
\;\;\frac{1}{2}\lambda^{\prime\prime}  v_1 v_3 +\nu_{13}^2,\\
&  -\frac{\lambda^{\prime\prime}v_2(v^2_1+v^2_3)+2(v_1\nu^2_{12}+v_3\nu^2_{23})}{2v_2} &
\;\;\frac{1}{2}\lambda^{\prime\prime}   v_2 v_3+ \nu^2_{23}  \\
& & -\frac{\lambda^{\prime\prime}v_3(v^2_1+v^2_2)+2(v_1\nu^2_{13}+v_2\nu^2_{23})}{2v_3}
\end{array}\right).
\label{mcharged}
\end{equation}
With the respective mass eigenvectors being denoted by $(H^+_{1,2,3})$. The mass matrix in Eq.~(\ref{mcharged})
has $\textrm{Det} M^2_c=0$ and hence this sector has a charged Goldstone boson.

Below we will find realistic mass spectra by diagonalizing the mass matrices given in (\ref{mpseudo}), (\ref{mfisicos})
and (\ref{mcharged}). We assume also that all VEVs are positive.

Consider the case when $\mu^2_i\not=\nu^2_{ij}\not=0,\;i,j=1,2,3;i\not=j$ and
$\mu^2\gg \nu^2_{ij},\,\forall i,j$. For the sake of simplicity we assume also that $v_1\ll v_2=v_3=v$ and dubbed $r=v_1/v\ll1$
and $\nu^2_{12}=\nu^2_{13}$.
Under these conditions the pseudoscalar masses from Eq.~(\ref{mpseudo}), are :
\begin{eqnarray} &&m^2_{a1}= 0, \nonumber \\&&
 m^2_{a2}= - 2 \lambda_3 (2  + r^2)v^2-\left(\frac{2}{r}+r\right)\nu^2_{13},\nonumber\\&&
 m^2_{a3}= -2\lambda_3(2+r^2)v^2-r\nu^2_{13}-2\nu^2_{23},
\label{mp2}
\end{eqnarray}
and $\nu^2_{13},\nu^2_{23}<0$ and the sign of $\lambda_3$ depend on the magnitude of the soft terms.
There is one Goldstone boson as it must be. All the real neutral scalar masses are non-zero:
\begin{eqnarray}
&&m^2_{s1} =  (2\lambda-\lambda^\prime) v^2-2 \nu^2_{23}-r \nu^2_{13},
\nonumber \\ && m^2_{s2}=A +\frac{1}{2r} \sqrt{B},\; m^2_{s3}=A -\frac{1}{2r}\sqrt{B},
\label{ms2}
\end{eqnarray}
with $A$ and $B$ defined as:
\begin{eqnarray}
&& A=\left[\lambda(1+r^2)+\frac{\lambda^\prime}{2}\right]v^2-\left(\frac{1}{r}+\frac{r}{2} \right)\nu^2_{13}
\nonumber \\ && B=[(2\lambda r(1+r^2)+\lambda^\prime r)v^2-(2+r^2)\nu^2_{13} ]^2\nonumber \\&&
+8r[(\lambda(\lambda^\prime-2\lambda)+\lambda^{\prime\,2})r^3v^4+(2\lambda+r^4+\lambda^\prime(1+2r^2))v^2\nu^2_{13}]
\label{ab}
\end{eqnarray}
Finally, the masses for the charged scalars from Eq.~(\ref{mcharged}), are
\begin{eqnarray}
&&m^2_{c1}= 0, \;m^2_{c2}=-\frac{1}{2}\left( \frac{2}{r}+r\right)(\lambda^{\prime\prime}rv^2+2\nu^2_{13})\nonumber \\ &&
m^2_{c3}=-\frac{\lambda^{\prime\prime}}{2}(2+r^2)v^2-r\nu^2_{13}-2\nu^2_{23},
\label{mc2}
\end{eqnarray}
in which we have also only one Goldstone boson, as it must be.

For instance, if the triplet is one of those that contribute to the neutrino masses in \cite{dim5}, i.e.,
$v\sim10^{-3}$ GeV and $v_1\sim10^{-7}$ GeV, then $r=v_1/v=10^{-4}\ll1$~\cite{dim5}.
Notice that, since $\nu$'s $\gg v,v_1$ and also $r$ is small, the hierarchy among the scalar masses does not imply
a respective hierarchy among the $\lambda$s, i.e., they may
be of the same order of magnitude. Expanding in $r$ (i..e, $r\ll 1$) we obtain from (\ref{mp2})--(\ref{mc2}):
\begin{eqnarray}
&&m^2_{a2}\approx -\frac{2}{r}\nu^2_{13},\quad m^2_{a3}\approx-2\nu^2_{23}-r\nu^2_{13}
\nonumber \\&&
m^2_{s1}\approx-2\nu^2_{23}-r\nu^2_{13},\quad m^2_{s2}\approx (2\lambda+\lambda^\prime)v^2 ,
\quad m^2_{s3}\approx -\frac{2}{r}\nu^2_{13},\nonumber\\&&
m^2_{c2}\approx-\frac{2}{r}\nu^2_{13},\quad m^2_{c3}\approx -2\nu^2_{23}-r\nu^2_{13}.
\label{oba2}
\end{eqnarray}
From (\ref{oba2}) we see that all pseudoscalar and charged Higgs bosons may be heavy even for
small  VEVs if the $\nu^2$s are large enough (and negative). Then, when all VEVs are small, i.e., of the order of
MeV or less, there is a light real neutral scalar, $s_2$.

This is not an artifact of the approximation used in obtaining the square masses in (\ref{mp2}), (\ref{ms2}) and
(\ref{mc2}). We have obtained numerically the eigenvalues of the full matrices (\ref{mpseudo})-(\ref{mcharged})
i.e., without assuming $v_2=v_3$ and $\nu_{12}=\nu_{13}$, and found that a light real scalar always survives.
Just as an example, let us consider one of the triplets $H=(H_1,H_2,H_3)$ related to
the neutrino mass generation in the model of Ref.~\cite{dim5}.
For simplicity we assume that the dimensionless coupling constants are $\lambda_1=\lambda_2=\lambda_3=\lambda_4=1$. The
VEVs are (in GeV) $v_1=0.000001$, $v_2=0.00002$ and $v_3=0.001$, and the soft parameters (in GeV$^2$) $\nu^2_{12}=-100$,
$\nu^2_{13}=-200$ and $\nu^2_{23}=-300$, we obtain the masses for the physical scalars (in GeV): $m_{a1}$ and $m_{c1}$
vanish, they are the Goldstone bosons,
$m_{s1}\sim m_{c2}\sim m_{a2}\approx122.5$, $m_{s2}\approx 2\times10^{-3}$, and $m_{s3}\sim m_{c3}\sim m_{a3}\approx450$.
Notice that $H^0_1$, with mass $m_{s1}$ may be the neutral scalar
which correspond to the SM Higgs scalar, $H_{SM}$. However, in this case it couples mainly to neutrinos.
For the VEVs of the other doublets, denoting the respective masses as $(m_{s1},m_{s2},m_{s3})$ we obtain
(in GeV) $(65,84,466), (5.7,25,302),(116,244,5895),(98,137,230)$ for the VEVs of the  $H^{\prime\prime}
(\Phi^{\prime\prime}),(H^\prime),(\Phi^\prime)$,  in the notation of \cite{dim5}, respectively. The triplet with the
lightest neutral scalar belong to $\Phi^{\prime\prime}$, notwithstanding, their coupling with $d$-like quarks
are suppressed by the factor $1/\Lambda$, see Eq.~(4) in Ref.~\cite{dim5}.

One of the triplets of doublets has, at the tree level,
a light neutral scalar, with mass of the order of the MeV, see $m_{s2}\approx 2$ MeV above.
This sort of scalar boson can be discovery by the shift in the energy level in muonic
atoms~\cite{jackiw}. If  the shift caused by scalar exchange is less than the experimental value~\cite{aas}
a scalar mass $\leq 9$ MeV is ruled out~\cite{borie}. This limit however may be evaded in the case of multi Higgs
model and lighter scalar may be allowed because there are mixing among the scalars, see Ref.~\cite{qmasses}.
It happens also that loop corrections must rise the
mass $m_{s2}$ to a value above this limit (see below). We must stress that usually the search for light scalars in nuclear
and rare meson decays is based on the assumption that the scalar is the one of the SM~\cite{higgshunter}.

On the other hand, if all $\nu$s vanish  and all the VEVs in Eqs.~(\ref{mpseudo})--(\ref{mcharged}) are of the order of
several GeVs, say $\sim 174/\sqrt{3}$ GeV, all Higgs scalars are heavy as well. Notwithstanding,  this case is not interesting
in the context of the models of \cite{dim5} and \cite{porto} if neutrino masses are included in the latter model.

\subsection{Third case: model with no soft $A_4$ violating terms}
\label{subsec:caso3}

The case when  $\mu^2_i=\nu^2_{ij}=0,\;i,j=1,2,3;i\not=j$ admits two type of solutions: i) $v_1=v_2=v_3$ and
ii) $v_1 \not= v_2 \not=v_3$ (and all of them different from zero).
In the case i), the relations in (\ref{minimo}) are reduced to only one. In this case all mass
matrices have all the same form: non-diagonal elements are all equal (denoted by $b$) and the  diagonal
elements are also equal to each other (denoted by $a$). The eigenvalues are $a+2b$ and $a-b$ and $a-b$
(the values of $a$ and $b$ depend on the sector: real, imaginary and charged sector).
All real scalars are massive, one on them may, or may not, corresponds to the SM Higgs doublet and the other two
are mass degenerated. This case was considered in detail in Ref.~\cite{ema}, see also Ref.~\cite{zee}.

The case ii) is consistent with the simultaneous solution of the constraint equations Eq.~(\ref{minimo}),
$t_i=0$, only if we have $3\lambda_2=2\lambda_3+\lambda_4$ [i.e.,
$-\mu^2=(\lambda_1+\lambda_2)V^2$ where $V^2=v_1^2+v_2^2+v_3^2$] and $\lambda_1+\lambda_2>0$.
The corresponding mass spectrum for the pseudoscalars is:
\begin{equation}
m^2_{a1}=0, \quad m^2_{a2}=m^2_{a3}=-2\lambda_3V^2,
\label{ps0}
\end{equation}
while in the neutral real scalars we have
\begin{equation}
m^2_{s1}=m^2_{s2}=0,\quad m^2_{s3}=
2(\lambda_1+\lambda_2)V^2.
\label{s0}
\end{equation}
In the charged scalars sector we find:
\begin{equation}
 m^2_{c1}=0,\quad m^2_{c2}=m^2_{c3}=-\frac{3}{2}\lambda_2V^2.
\label{cs0}
\end{equation}

From  the  above  results we have the following constraints on the
scalar potential parameters: $\lambda_{2,3}<0$, $\lambda_1+\lambda_2>0$ and
$\lambda_2=(2\lambda_3+\lambda_4)/3$, which can be  satisfied provide
\begin{equation}
\lambda_4>0,\quad\lambda_3<-\frac{\lambda_4}{2},\quad
-\lambda_1<\lambda_2<0.
\label{cond2}
\end{equation}

At this level, this situation is unrealistic since it has two massless neutral boson at
the tree level, thus at this level it can be considered ruled out
unless the coupling of these scalars and light fermions are suppressed.
However, we must stress that in the model of \cite{dim5}
it is still possible to avoid massless scalars at the tree level since in the model
there are trilinear interactions involving two $A_4$ triplets and one singlet of $SU(2)$ and $A_4$.
Denoting the second triplet of doublets by $\Phi$ with  large VEVs and a scalar singlet ($Y=0$),
$\zeta$, the $A_4$ symmetry may allow trilinear interactions like $f[h\Phi]_1\zeta$, $f$ is an arbitrary
energy scale of the order of the electroweak scale. In this case, the masses related to $h$ are risen up
and there is no light real neutral scalars. On the other hand, since there is no symmetry  protecting
these zero masses fields, small non-zero masses can be generated by radiative corrections. See below.

\section{Conclusions}
\label{sec:con}
\bigskip
The third case with the solution ii) with just three doublets is not realistic at the tree level.
However, as we said before, in the context of models with other doublets and singlets with large VEVs the light
scalar becomes heavy by quantum corrections. It is well known that if there exist a symmetry protecting the
mass of a particle, say $m^2$, the corrections to $m^2$ are of the order of $m^2$. However, if that  symmetry
does not exist, nothing protects $m^2$ from become as large as the energy scale where new physic may arise.
In particular, scalar masses are difficult to keep smaller than the higher energy scale until which
the model is renormalizable~\cite{veltman81}.
In this vain, we can make some general considerations. \textit{1)}
For instance if the light scalar say $H_l$, with mass $m_l$, couples with a heavy field $H_h$ of mass $M_h$, and
with the coupling constant $\lambda$ between them, say $\lambda H^2_lH^2_h$,
or if there exist a trilinear interaction involving a singlet $\zeta$, say $fH_h H_l\zeta$,
at the 1-loop level we can only say that, assuming that there is no
unexpected cancelations the mass of the light scalar gets quantum corrections $m_l<M_h/4\pi$ or $m_l<f/4\pi$~\cite{burguess}.
Model like those in \cite{dim5,porto} have the hierarchy problem as any multi-Higgs extension of the SM.
However, once this problem is solve by any mechanism, say extra dimensions, and the heaviest scalar is stable against
quantum corrections, this mass becomes a cut off for the lighter scalars, i.e., it is an upper bound for the masses of
these scalars. The mixing patterns also do not depend on fine-tuned and \textit{ad-hoc} values of the parameters but are
determined, like masses, by the vacuum structure. \textit{2)} The stability of the vacuum sets a lower bound on the scalar
mass but it is model dependent. In the context of the SM it is about 7 GeV~\cite{higgshunter}.
This lower limit depend on the quark top mass and, with the known value of this mass, $m_t\sim172$ GeV, it is not possible
to have a light neutral scalar boson anymore. However, in multi-Higgs models with no scalar singlets this
lower bound applies to the heaviest Higgs boson mass~\cite{weinberg}, thus in those models a very light scalar boson
may be consistent with the present value of $m_t$.
\textit{3)} In realistic models, as that of Ref.~\cite{dim5}, beside the mechanism discussed in \textit{1)},
it is possible that the Higgs scalars having small VEVs interact mainly with neutrinos and besides
these interactions may be only through non-renormalizable effective operators that are
suppressed by power of the cut off $\Lambda$ which is of the order of TeV, as is the case of the model of \cite{dim5}.
\textit{4)} Last but not least, notice that in the second case  the pseudoscalar physical states are heavy enough to
not be produced together with its real partner in $Z$ decay thus, there is no contribution to the invisible $Z$-decay width,
even when all the VEVs are small.

In summary, we have shown that under some condition, in a model with three scalar doublets with an $A_4$ symmetry
the scalar mass spectra may have no light scalar even if some of the VEVs
are smaller than the electroweak scale. In cases were light scalars are unavoidable at the tree level,
these fields may be safe from the phenomenological point of view since they couple mainly to neutrinos
and/or they become heaviest due to quantum corrections. The large hierarchy among the fermion masses are
now a hierarchy in the VEV structure i.e., of the vacuum whatever dynamics underlies this mechanism.
This mechanisms have been implemented using only three $SU(2)$ scalar doublets, but in more realistics models
involving more doublets, triplets and singlets it will work as well.

A similar model with 3 Higgs doublets and $A_4$ symmetry
has been studied \cite{3alianca}. The difference between our work and theirs are the following: i) we have considered
no CP violation,  ii) we are in the context of the model of \cite{dim5} in which all VEVs have to be different from zero
and for these reason we have omitted the possibilities in which one or two VEVs are zero; iii) the phenomenology of
our model is being considered and submitted for publication as soon as possible\cite{qmasses}.


\end{document}